\documentclass[10pt,sigconf]{acmart}
\usepackage{array}
\usepackage[english]{babel}
\usepackage[utf8]{inputenc}
\usepackage{blindtext}
\usepackage{threeparttable}
\usepackage{xcolor}
\usepackage{lmodern}
\usepackage{tabularx}
\usepackage{colortbl}

\usepackage{tikz}

\newcommand\encircled[1]{%
  \tikz[baseline=(X.base)] 
    \node (X) [draw, shape=circle, inner sep=0, fill=black, text=white] {\strut #1};%
}

\usepackage{balance}

\usepackage{enumitem}
\setlist{nolistsep}
\usepackage{graphicx}
\usepackage{epstopdf}
\usepackage{grffile}
\usepackage{amsmath}
\usepackage{multicol,multirow}
\usepackage{rotating,url}
\usepackage{enumitem}
\usepackage{color}
\usepackage{subcaption}
\usepackage{xspace}
\usepackage{ifpdf}
\usepackage{mdwlist}
\usepackage{colortbl}
\usepackage{booktabs}
\usepackage{multirow}
\usepackage{enumitem}
\usepackage{xspace}
\usepackage{cancel}
\usepackage{xfrac}
\usepackage[absolute,showboxes]{textpos}
\usepackage{afterpage}% http://ctan.org/pkg/afterpage
\newlength{\oldtextfloatsep}
\usepackage{cleveref}

\newcommand{\eat}[1]{}
\usepackage{tabularx}
\newcolumntype{L}[1]{>{\raggedright\arraybackslash}p{#1}} % linksbündig mit Breitenangabe
\newcolumntype{C}[1]{>{\centering\arraybackslash}p{#1}} % zentriert mit Breitenangabe
\newcolumntype{R}[1]{>{\raggedleft\arraybackslash}p{#1}} % rechtsbündig mit Breitenangabe
\long\def\comment#1{}

\begin{document}

\title[]{A Tale of Two Markets: Investigating the Ransomware Payments Economy}

%\titlenote{Produces the permission block, and copyright information}
%\subtitle{Extended Abstract}

\author{Kris Oosthoek}
\affiliation{
   \institution{Delft University of Technology}
}

\author{Jack Cable}
\affiliation{
   \institution{Independent Researcher}
}

\author{Georgios Smaragdakis}
\affiliation{
   \institution{Delft University of Technology}
}

%\renewcommand{\shortauthors}{XXX et al.} %{X.et al.}

% Disable / remove copyright boxes -- Do not move it upwards
\setcopyright{none}
\settopmatter{printacmref=false, printccs=false, printfolios=false}
\renewcommand\footnotetextcopyrightpermission[1]{}
\pagestyle{plain}
\acmConference{}{}{}
\renewcommand{\shortauthors}{}

\begin{abstract}

Ransomware attacks are among the most severe cyber threats. They have made headlines in recent years by threatening the operation of
governments, critical infrastructure, and corporations. Collecting and
analyzing ransomware data is an important step towards understanding the spread
of ransomware and designing effective defense and mitigation mechanisms.  We
report on our experience operating {\tt Ransomwhere}, an open crowdsourced
ransomware payment tracker to collect information from victims of ransomware
attacks. With {\tt Ransomwhere}, we have gathered 13.5k ransom
payments to more than 87 ransomware criminal actors with total payments of more
than \$101 million. Leveraging the transparent nature of Bitcoin, the cryptocurrency used for
most ransomware payments, we characterize the evolving
ransomware criminal structure and ransom laundering strategies. Our analysis shows that there are two parallel ransomware
criminal markets: commodity ransomware and Ransomware as a Service (RaaS).
We notice that there are striking differences
between the two markets in the way that cryptocurrency resources are utilized,
revenue per transaction, and ransom laundering efficiency. Although it is
relatively easy to identify choke points in commodity ransomware payment activity, it
is more difficult to do the same for RaaS.

%, the latter which became dominant after 2019. 
%\todo{@George: rewrite} 
%Ransomware is a form of malware designed to unauthorizedly encrypt files on a
%third-party device and make them unusable. 
%In its cyber-criminal form, as
%Ransomware as a Service (RaaS) it is a major concern of CISOs and any team with
%a cyber security responsibility. 
%Ransom payments of several million USD, generally 
%in Bitcoin, have become a regularity. Apart from devastating impacts to operations
%and finance, with RaaS organizations, institutions and governments also face the
%threat of leakage of sensitive data due to double extortion by these actor groups.
%
%In this paper, we analyze a dataset of 7,321 Bitcoin addresses associated with more
%than 13,497 ransomware payments in the last five years. Our analysis on associated
%transactions shows that
%ransomware criminal activity has evolved over the years. Low-value ransom
%incidents provide payment methods and addresses in the cryptocurrency systems in
%plain text. On the other hand, high-profile ransom incidents unveil their
%payment address only after negotiations. The criminal organizations behind this
%business type differ than those in the low-value ransom and form a new market:
%ransomware as a service (RaaS). We notice that the payment address registration
%and laundering of ransom in RaaS deviate substantially compared to common
%practices in low ransome incidents.  We characterize the two markets and
%investigate ways to defend against these types of attacks by understanding the
%structure and operation of the two markets.

\end{abstract}

\maketitle

\section{Introduction}\label{sec:intro} %\todo{@George: first draft;1 page for abstract+intro}

Ransomware, a form of malware designed to encrypt a victim's files and make them
unusable without payment, has quickly become a threat to the functioning of many
institutions and corporations around the globe. In 2021 alone, ransomware caused
major hospital disruptions in Ireland~\cite{NHS-ransomware}, empty supermarket
shelves in the Netherlands~\cite{Cheese-ransomware}, the closing of 800
supermarket stores in Sweden~\cite{Coop-Sweden-ransomware}, and gasoline
shortages in the United States~\cite{Gasoline-ransomware}. In a recent report,
the European Union Agency for Cybersecurity (ENISA) ranked ransomware as the
``prime threat for 2020-2021''~\cite{ENISA-threat2021}. The U.S. government
reacted to high profile attacks against U.S. industries by declaring ransomware a
national security threat and announcing a 
``coordinated campaign to counter ransomware''~\cite{NSA-Year-In-Review}. Other governments,
including the United Kingdom~\cite{UK-ransomware}, Australia~\cite{Australia-ransomware},
Canada~\cite{Canada-ransomware}, and law enforcement agencies, such as the
FBI~\cite{FBI-ransomware} and Europol~\cite{Europol-ransomware}, have also
launched similar programs to defend against ransomware and offer help to
victims.

To the criminal actors behind these attacks, the resulting disruption is just
`collateral damage'. %They exploit with financial intent. And with success. 
A handful of groups and individuals, with names such as NetWalker, Conti, REvil
and DarkSide, have received tens of millions in USD as ransom.  But this is just
the top of the food chain in an ecosystem with many grey areas, especially when
it comes to laundering illicit proceedings. In this article, we will provide a
closer look at the ecosystem behind many of the attacks plaguing businesses and
societies, known as Ransomware as a Service (RaaS).

Cryptocurrencies remain the payment method of choice for criminal ransomware
actors.  While many cryptocurrencies exist, Bitcoin is preferred due to its
network effects, resulting in wide exchange options. Bitcoin's sound monetary
features as a medium of exchange, unit of account and store of value make it as
attractive to criminals as it is to regular citizens. According to the U.S.
Department of Treasury, based on data from the first half of 2021, the ``vast
majority'' of reported ransomware payments were made in Bitcoin~\cite{fincen}.
%Nonetheless, the Department notes a small number of transactions made in Monero,
%a privacy-preserving cryptocurrency that makes tracking payments much more difficult.
%%%%%% commented this out to decrease the Introduction a bit
Law enforcement agencies have started to disrupt ransomware actors by
obtaining personal information of users from Bitcoin exchange platforms. This is realized through
anti-money laundering regulations such as Know Your Customer (KYC), which require
legal identity verification during registration with the service. While
cryptocurrencies such as Bitcoin are enablers of ransomware, blockchain
technology also offers unprecedented opportunities for forensic analysis and
intelligence gathering.
Using our crowdsourced ransomware payment tracker, {\tt Ransomwhere}, we compile
a dataset of 7,321 Bitcoin addresses which received ransom payments, based on
which we shed light on the structure and state of the ransomware ecosystem. 

\noindent Our contributions are as follows: 

\begin{itemize} [leftmargin=*,topsep=0ex]

\item We collect and analyze the largest public dataset of ransomware activity to date,
which includes 13,497 ransom payments to 87 criminal actors over the last
five years, worth more than 101 million USD. 
%\todo{101 or 107 million?} 101 - a small difference due to the exclusion of some addresses, but perhaps more perhaps due to a difference in source for the daily close rate; e.g. CoinMarketCap is unreliable because these include platforms which wash trade for high volumes; I used CoinDesk, but this ends up some percentages lower than other sources

\item We characterize the evolving ransomware ecosystem. Our analysis shows that
there are two parallel ransomware markets: %, namely, 
commodity and 
%Ransomware as a Service (RaaS).
RaaS. 
After 2019, we observe the rapid rise of RaaS, which achieves higher revenue per address and
transaction, and higher overall revenue.
%distinct periods, namely, before and after 2019. 

\item We also characterize ransom laundering strategies by commodity ransomware and
RaaS actors. Our analysis of more than 13k transfers shows striking differences
in laundering time, utilization of exchanges and other means to cash out ransom
payments.

\item We discuss the difficulties defending against professionally operated
RaaS and we propose possible ways to trace back RaaS activity in cryptocurrency
systems.

\item To enable future research in this area, we make our
tracker, {\tt Ransomwhere}, and the tracked ransomware payments of our analysis
publicly available at~\cite{Ransomwhere}.

\end{itemize}

%- We characterize the RaaS ecosystem by classifying victims based on
%  announcements by these groups on their darkweb pages.

%- We estimate the revenues of X RaaS groups.

%- We track the techniques applied by these groups in laundering Bitcoin to fiat
%  off-ramps.

%\textbf{Space allocation as Introduction will probably take 1 page} \newpage

\section{The Ransomware Ecosystem}\label{sec:ecosystem} %\todo{1 page}

The ransomware ecosystem and its payment traffic can be largely recognized
in two categories: commodity ransomware and ransomware as a service (RaaS). 
%We discuss the characteristics of each in this section.

\subsection{Commodity Ransomware}\label{sec:commodity}

%Up until 2019, %let's keep the 2019 mark as a result of our analysis
In the early years of ransomware, the majority of ransomware that spread can
be characterized as `commodity' ransomware. Commodity ransomware is
characterized by widespread targeting, fixed ransom demands, and
technically-adept operators. It usually targets a single device
\cite{kharraz2015cutting}. Actors behind commodity ransomware are usually
technically savvy, as most of the time it is developed and spread by the same
person. Commodity ransomware operators take advantage of preexisting work, often
copying and modifying leaked or shared source code, causing the formation
of ransomware \textit{families}. Historically, most commodity ransomware
campaigns utilized phishing emails as the primary delivery vector and exploited
vulnerabilities in common word processing and spreadsheet software, if not directly via malicious executables. The modus operandi was
mass exploitation, rather than targeting specific victims.

Exemplary are WannaCry and NotPetya ransomware families, which over a course of only two months
impacted tens of thousands of organizations in over 150 countries by exploiting
a vulnerability allegedly stolen from the NSA~\cite{notpetyawannacryguardian}.
In today's standards, both families were poorly coded and their payment systems
weren't ready for business, although allegedly on purpose for NotPetya
\cite{notpetya}. 
%A commercially successful commodity ransomware family is Locky.
%Its server-side generation of decryption keys rendered manual file decryption
%near impossible, which led to many victims giving in and paying a relatively
%small sum in order to access their files. It has many variants, which are spread
%by Necurs, one of the largest botnets to exist~\cite{lockynecurs}.

With regards to its mitigation, the conventional advice is to have a proper
backup and contingency plan. The initial philosophy was that a quick ability to
restore would make it unnecessary to pay, impairing the financial incentive of
ransomware operators. But it turned out that what we now regard a commodity was
just a proving ground for a higher-impact utilization of ransomware.

\subsection{Ransomware as a Service (RaaS)}\label{sec:raas}

While the first reports of Ransomware as a Service (RaaS) emerged in 2016, it
wasn't until 2019 that RaaS became widespread, rapidly capturing a large share
of the ransomware market. We define RaaS as ransomware created by a core team of
developers who license their malware on an affiliate basis. They often provide a
payment portal (typically over Tor, an anonymous web protocol), allowing
negotiation with victims and dynamic generation of payment
addresses (typically Bitcoin). RaaS frequently employs a double extortion scheme,  not only encrypting victims' data, but also threatening to leak their
data publicly.

% Rather than being distributed by the developer itself, RaaS is operated by
% collectives of cyber criminal actors.
% RaaS groups now account for the majority of
% today's revenue from ransomware.

% Where ransomware was once the sole territory of
% those with coding skills, now complete toolkits with malware builder modules,
% custom ransom notes and even customer support pages are offered in a franchise
% model. Launching a ransomware campaign has become as easy as ordering from
% Amazon.

The rise of RaaS has enabled existing criminal groups to shift to a new
lucrative business model where lower-skilled affiliates can access exploits and
techniques previously reserved for highly-skilled criminals. This was
exemplified by a leaked playbook from the RaaS group Conti, which enables novice
actors to compromise enterprise networks~\cite{contiplaybook}.  RaaS affiliates
can differ markedly in their approaches. Some scan the entire internet and
compromise any victims they can. Once they have identified the victim, they
engage in price discrimination based on the victim company's size.  Affiliates
may even use financial documents obtained in the attack to justify higher
prices~\cite{microsoftransomwarereport}. Another strategy, known as \textit{big game
hunting}, targets big corporations that can afford paying a high ransom.
Darkside is one of most notable RaaS families whose affiliates practice big game
hunting, including the notable Colonial Pipeline attack in 2021~\cite{darkside}.

% RaaS can also be regarded as ransomware employed as part of a broader attack
% campaign against a specific target. While commodity ransomware is spread widely,
% RaaS targets specific victims. It is placed by a human attacker with 'hands on
% the keyboard', who will usually attempt to exploit vulnerabilities in one or
% more Internet-facing devices before executing the ransomware. As opposed to
% commodity ransomware, its impact usually reaches further than a single system,
% but rather is enterprise-wide.

%REvil, etc.
%
%Run by groups

RaaS families often rely on spear-phishing over the mass phishing mails
utilized by commodity ransomware groups.  They also exploit recently disclosed
vulnerabilities, leaving remote and virtual desktop services
vulnerable~\cite{cisatopexploits}.  RaaS has lowered the barrier to entry into
cyber-criminality, as it has removed the initial expenditure to develop effective
ransomware. As a result, attacks can be performed with near zero cost.
Combined with high ransom demands, this has led to a low-risk, high reward
criminal scheme.

RaaS has effectively {\it weaponized} the unpatched internet-facing technology
of many unwitting organizations. Such organizations have significant financial
interest to have systems restored and get back to business after a ransomware
attack. Cryptocurrencies enable ransomware actors to directly monetize these
vulnerabilities at a scale never seen before. In this paper, we regard the
functioning of ransomware actors through what usually is the last mile of the
attack.

\begin{figure}[t]
	\includegraphics[width=\columnwidth]{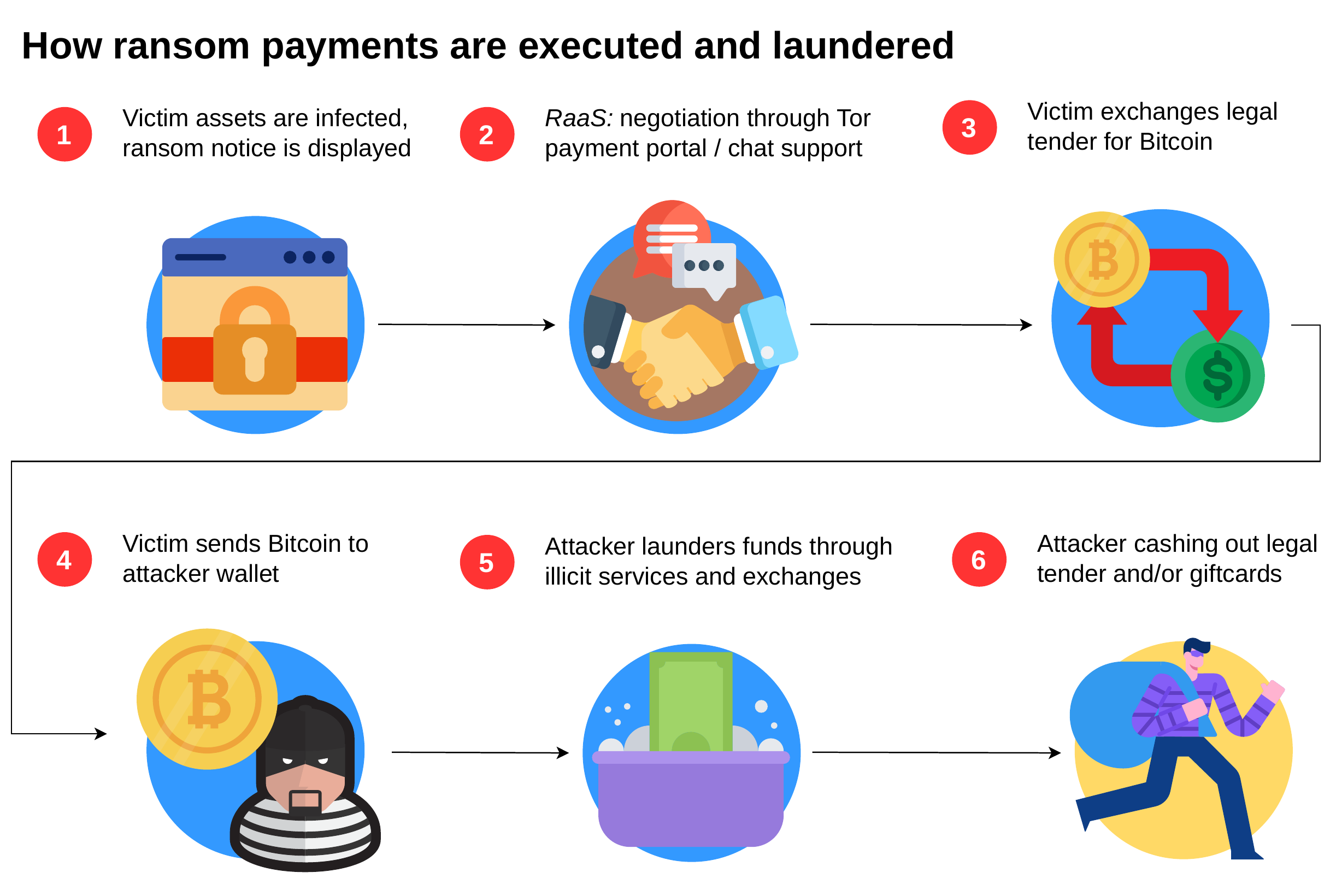}
	\caption{General course of a ransom payment and its laundering.}
	\label{fig:cartoon}
\end{figure}

Figure \ref{fig:cartoon} shows the general course of events after a ransomware infection, when the victim decides to pay the
attacker (step~\encircled{1}). In the case of commodity ransomware families,
the ransom demand price is fixed and negotiation with the attacker isn't necessary.
With RaaS, attackers usually run chat-based services to interact with victims and negotiate the final ransom amount (step~\encircled{2}). After this, a
victim will usually exchange fiat legal tender for cryptocurrency such as Bitcoin at an
exchange platform (step~\encircled{3}) and then send it to the attacker's wallet
(step~\encircled{4}). The attacker will then usually route the obtained Bitcoin
through various services (step~\encircled{5}) in order to obfuscate ownership and
reduce the risk of deanonymization before cashing out (step~\encircled{6}).

\section{Methodology}\label{sec:methodology}    %\todo{1 page}

In this section, we describe how we collected data of ransom payments and ransomware actors in
our study. 

%\subsection{Ransomware Dataset}\label{sec:datasets}

\subsection{Addresses involved in Ransom Payments}\label{sec:addresses}

We obtain ransomware Bitcoin addresses from our crowdsourced payment tracker
{\tt Ransomwhere}. The {\tt Ransomwhere} dataset contains Bitcoin addresses and
associated families collected from open-source datasets and publicly-submitted
crowd-sourced reports. In total, the Ransomwhere dataset contains 7,457 Bitcoin
addresses and their corresponding ransomware families.

To seed the dataset, we collected data from several public sources. We imported
addresses from Paquet-Clouston et al.~\cite{paquetclouston2018ransomware}, who
collected 7,222 addresses and labeled families representing approximately
\$12.7 million in payments. This dataset provides us with, among other
ransomware families, 7,014 addresses belonging to Locky.  We further collected
37 addresses and associated families from the AT\&T Alien Labs Open Threat
Exchange, an open threat intelligence sharing platform~\cite{alienvault}.  

Members of the public may submit reports at our crowdsourced payment
tracker {\tt Ransomwhere}~\cite{Ransomwhere}.  We received 99 reports containing 198 addresses over a
6-month period from June 2021 to December 2021. While this is a lower number of addresses, they represent the majority of ransomware payment value in our dataset. In order to verify reports, the
reporter must include the relevant Bitcoin addresses and the associated ransomware
family. In addition, they have to provide evidence of the ransom demand, such as a screenshot of the
ransom payment portal or a ransom message on an infected computer. Some
addresses were involved in more than one report. All reports were manually
reviewed before being added to the dataset. We did not accept reports that were
inaccurate or were not related to ransomware (e.g., addresses involved in
extortion scam emails).

All reported ransom addresses were Bitcoin addresses. Due to the transparent nature
of Bitcoin it is possible to verify that the collected addresses indeed received
payments. Using our own Bitcoin full node, we scraped all transactions for
the addresses in our dataset. Overall, 7,323 out of 7,454 Bitcoin addresses were
involved in at least one ransom payment. We discarded 134 addresses that did not receive any payment.
We have queried Tor using a solution from a peer researcher~\cite{DWS} for all
Bitcoin addresses in our dataset to rule out the chance of an address being used
for cybercrime purposes other than ransomware. Based on this, we excluded 2
addresses belonging to a cache of Bitcoin seized by the U.S. Department of
Justice after the closing of the SilkRoad darkweb market~\cite{Silkroad}. After 
these steps, the final number of addresses considered for our analysis is 7,321.
For a summary of our dataset we refer to Table~\ref{table:datasetoverview}. 

\subsection{Ransom Payments and Laundering}

The transparency of Bitcoin also allows us to collect information about (ransom)
payments, i.e., the amount of Bitcoin received. For each address we collected the
number of incoming (payments) and outgoing (transfers) transactions, their value in Bitcoin, and their
timestamp. We calculated the USD value of each transaction using the BTC-USD
daily closing rate on the day of the transaction. This serves as an approximate ransom payment and
not the exact amount in USD the criminal actors requested or later profited. The total ransom paid to addresses in our dataset is \$101,297,569. The lowest payment received is \$1, and the highest is \$11,042,163. The median payment value is \$1,176.

In collaboration with Crystal Blockchain~\cite{Crystal}, we tracked the
destination of outgoing transactions, i.e., transfers. In order to estimate addresses' potential
for illicit use, Crystal Blockchain utilizes clustering heuristics such as
one-time change address and common-input-ownership \cite{Bitcoinclustering}, as
well as human collection of off-chain data from various cryptocurrency services.
%\todo{I rephrased this sentence, is this a correct understanding?}.; yes it is
In addition to this, Crystal Blockchain scrapes online forums and other
Internet services for Bitcoin addresses and their associated
real-world entity. Based on this, it is possible to track payments several hops
from the original deposit address. To have the most reliable view, in
our analysis we have only regarded the direct destination of ransom payments
(first hop). Based on the characterization of the involved addresses across the
path, we are able to study the laundering strategies of ransomware groups as
well as the time needed to wash out the money (see
Section~\ref{sec:laundering}).

\subsection{Ransomware Actors}

We obtained addresses and labeled families as described in
Section~\ref{sec:addresses}. We categorized each ransomware family as used by
either commodity ransomware or RaaS actors. Ransomware is generally categorized as
RaaS due to the use of an affiliate structure, with the ransomware developer
(operator) selling the ransomware to criminal actors either based on a
commission for each ransom paid, or a flat monthly fee (\textit{as a service},
like many subscription-based services). As there does not exist any
comprehensive public list of RaaS groups, we have labeled a family as RaaS if
%(i) 
a reliable industry or law enforcement source claims that a given ransomware is sold \textit{as a service}. 
%(ii) when a Tor site used for double extortion or victim shaming is or has been
%available (verified using our Tor solution~\cite{DWS}) \todo{I'd recommend
%removing the second requirement and simply saying that this is a common
%charactaristic (e.g., some groups don't do double extortion but are still
%Raas)}. 
%When these criteria aren't met, the family was classified as commodity
%ransomware. 
A list of commodity and RaaS families in our dataset
is presented in Table~\ref{table:families}.

%Section X provides a full overview of the RaaS in our dataset.

%We list Ransomware as a
%Service families in Appendix A~\todo{list families}. 

\begin{table}[t]
	\caption{Ransomware Dataset Statistics}
	\scalebox{0.85}{
	\begin{tabular}{llll}
	\hline
	\textbf{Data}                 & \textbf{Commodity}  & \textbf{RaaS} & \textbf{Total} \\ \hline
	Unique Actors        & 71   & 16                           & 87             \\ \hline
	Bitcoin Addresses                      & 161 & 7,160           & 7,321                     \\ \hline
	Received Transactions & 4,799 & 8,698           & 13,497                    \\ 
	(Payments) &  &            &                     \\ \hline
	Transferred Transactions      & 4,557  & 8,540           & 13,097                       \\ 
	(Laundering)     &   &            &                        \\ \hline
\end{tabular} }%scalebox
	\label{table:datasetoverview}
\end{table}

\subsection{Limitations}

Our dataset of Bitcoin addresses is the largest public collection of ransomware
payment addresses collected to date, based on total USD value. While this allows
for a unique view on the ransomware financial ecosystem, it is not exhaustive.
An inherent limitation of any research using adversary artifacts is its
dependence on the availability of artifacts that bad actors have an interest to
hide. Furthermore victims might have an interest not to report addresses, as
they prefer keeping attacks undisclosed. We note that certain families, such as
NetWalker, may be overrepresented in our dataset due to us having more complete
data on these families.  Despite this limitation, we believe that our dataset
provides a valuable, if incomplete, representation of ransomware payments over
many years.  This broad view provides a better reflection of the state of
affairs than simply focusing on a few families. We hope that this can lay the
groundwork for further public data collection in the future, and encourage
anyone to submit data at {\tt Ransomwhere}~\cite{Ransomwhere}. %\href{https://ransomwhe.re}{ransomwhe.re}.

\section{Ransom Payment Analysis} %\todo{@George: confirm flow; 2.25 pages}

%In this section we will demonstrate that:
%
%- RaaS higher payments
%- RaaS less transactions per address, commodity more per address

\begin{table}[!t]
	\scalebox{0.7}{
	\begin{threeparttable}
	\centering
	\caption{Ransomware criminal actors in our Dataset} 
	\label{table:families}
		\begin{tabular}{ll|ll}
		\hline
		Name & \#Addrs. & Name (contd.) & \#Addrs. \\ \hline
		\cellcolor[gray]{.9}Locky & \cellcolor[gray]{.9}7037 & \cellcolor[gray]{.9}DarkSide & \cellcolor[gray]{.9}3 \\
		\cellcolor[gray]{.9}NetWalker & \cellcolor[gray]{.9}66 & \cellcolor[gray]{.9}MedusaLocker & \cellcolor[gray]{.9}3 \\
		SamSam & 48 & NotPetya & 3 \\
		\cellcolor[gray]{.9}Ryuk & \cellcolor[gray]{.9}40 & GlobeImposter & 3 \\
		\cellcolor[gray]{.9}Conti & \cellcolor[gray]{.9}27 & ThunderCrypt & 3 \\
		Qlocker & 22 & Nemucod & 3 \\
		JigSaw & 11 & \cellcolor[gray]{.9}LockBit 2.0 & \cellcolor[gray]{.9}2 \\
		CryptConsole & 10 & Globe v2 & 2 \\
		\cellcolor[gray]{.9}Egregor & \cellcolor[gray]{.9}9 & EDA2 & 2 \\
		DMALocker v3 & 9 & Flyper & 2 \\
		Globe v3 & 7 & Black Kingdom & 2 \\
		\cellcolor[gray]{.9}REvil & \cellcolor[gray]{.9}7 & CryptoLocker & 2 \\
		CryptoTorLocker2015 & 7 & \cellcolor[gray]{.9}AvosLocker & \cellcolor[gray]{.9}2 \\
		HC6/HC7 & 6 & NoobCrypt & 2 \\
		Globe & 5 & VenusLocker & 2 \\
		WannaCry & 5 & XLocker v5 & 2 \\
		TeslaCrypt & 5 & Chimera & 2 \\
		CTB-Locker & 5 & Badblock & 2 \\
		Xorist & 4 & Other Groups/Families* & 50 \\
        \hline
	\end{tabular}
	\begin{tablenotes}
		\small
		\item * 50 families with 1 address each. RaaS actors are highlighted.
	\end{tablenotes}
	\end{threeparttable}
	} %scalebox
\end{table}

In this section, we analyze 13,497 payments to the Bitcoin addresses owned by ransomware
actors in our dataset (see Table~\ref{table:datasetoverview}). A payment is a transaction received by an address
in our dataset. %Each transaction is regarded as a payment by a single victim. 
Table \ref{table:families} lists the ransomware families used by the actors in our
dataset. Our dataset contains Bitcoin addresses associated with 87 commodity
ransomware or RaaS actors. For reasons of brevity, families for which our dataset
contains just 1 address are excluded from Table \ref{table:families}. The 16
actors that are classified as RaaS, highlighted in Table \ref{table:families}, account for 7,160 out of 7,321 addresses in our dataset.
As mentioned previously, for
full review our dataset is publicly available~\cite{Ransomwhere}.

%Bitcoin transactions enable sending any amount of Bitcoin from one wallet
%address to another. Incoming Bitcoin transactions are effectively received
%transaction outputs. Outgoing transactions are known as spent outputs, while any
%Bitcoin remaining at a wallet address is referred to as unspent output.
Ransomware victims typically create an account with a reputable exchange
platform to buy Bitcoin with fiat currency. Then, victims perform a transaction
(payment) to the address provided by the ransomware actor. 
%This is supported by our blockchain analysis data. 
In our dataset, payment transactions to ransomware addresses tend to originate one to two hops away from reputable exchange platforms, such as Coinbase and
Kraken.

\begin{figure}
	\includegraphics[width=\columnwidth]{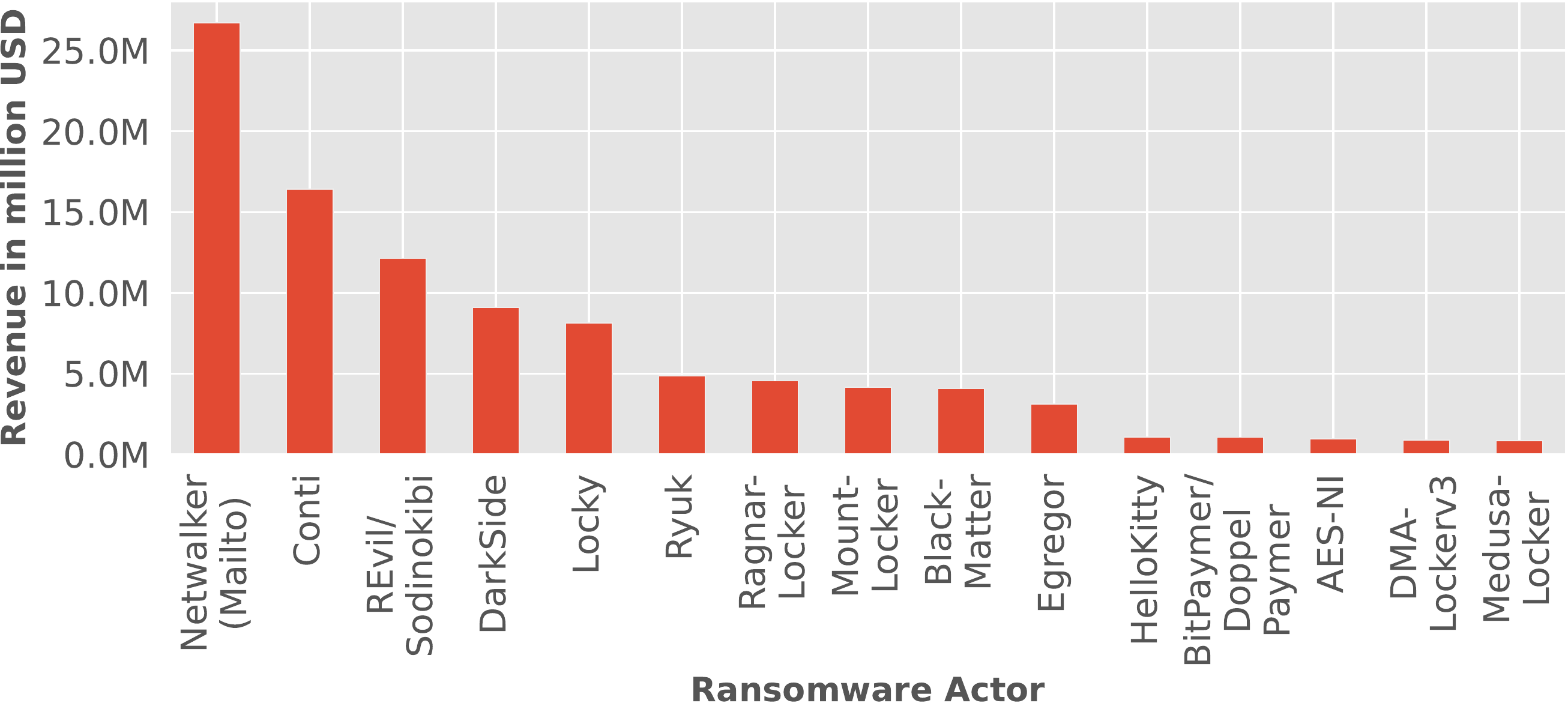}
	\caption{Revenue per ransomware actor.}
	\label{fig:rev-per-family}
\end{figure}

\subsection{Ransomware Revenue}

In Figure~\ref{fig:rev-per-family} we list 15 ransomware families with the
highest revenue. The top-grossing families are dominated by RaaS: NetWalker has
the highest revenue, \$26.7 million, followed by Conti (\$16.4 million),
REvil/Sodinokibi (\$12.1 million), DarkSide (\$9.1 million) and Locky (\$8.1 million).
All commodity actors combined account for a total revenue of \$5.5 million.  Although the number of
RaaS actors is lower, they together earned \$95.7 million. 

\begin{figure}
	\includegraphics[width=3.5in]{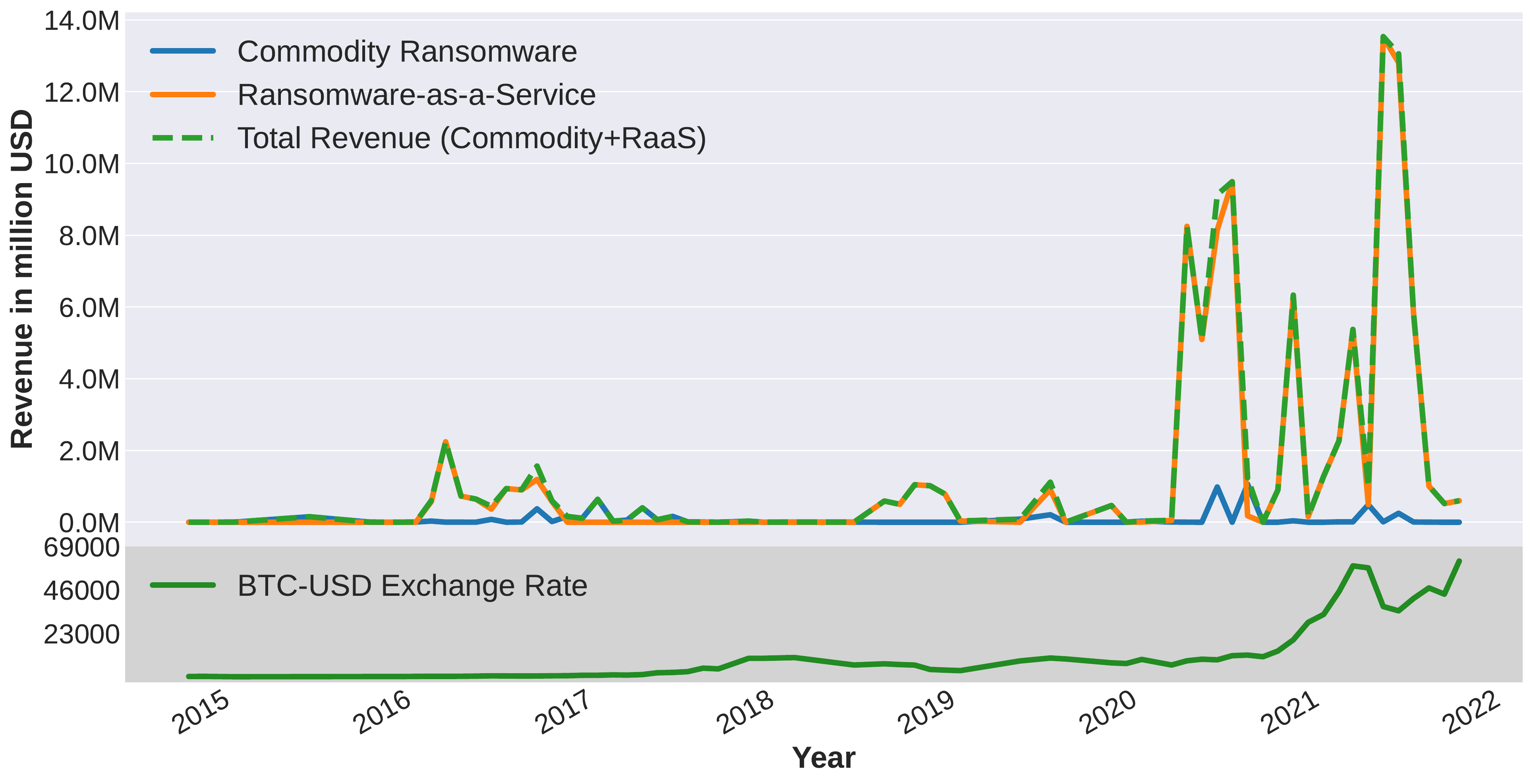}
	\caption{USD revenue for commodity and RaaS.}
	\label{fig:revenueperyear}
	\end{figure}

Figure \ref{fig:revenueperyear} shows the accumulated revenue of both commodity
ransomware and RaaS actors. It shows that, from 2015 until 2019, early RaaS
actors, primarily Locky, were gaining significant but still relatively low revenues.
Commodity actors were also active, but with even lower revenues.  As seen in
Figure \ref{fig:revenueperyear}, RaaS revenue reached \$8.2 million in April
2020.  This can be primarily attributed to NetWalker, which actively targeted
hospitals and health care institutions during the first COVID-19 lockdown in
that period~\cite{Netwalker}. Other revenue peaks caused by RaaS groups are in
May and June of 2021, with peaks of \$13.5 million and \$12.8 million
respectively.  These spikes are caused by large ransom payments by individual
victims. One example of this is a payment to REvil/Sodinokibi on June 1st, 2021,
accounting for \$11 million. This is a payment by the Brazilian meat processing
company JBS, which dominated headlines at the time \cite{JBSmeat}.

%The RaaS market is notable as it accounts for a much higher revenue than the commodity ransomware market, and in a shorter period of time.

Locky has a notorious reputation as one of the biggest ransomware strains in 2016-2017. It's also one of the earliest, if not the first, RaaS family. What stands out apart from its high revenue is its address usage. The actors behind Locky issued addresses to each new victim, a novelty at the time. This is evident in our analysis, with many addresses with only 2 or 3 incoming transactions.
%According to French court documents, the founder of the sanctioned
%cryptocurrency exchange BTC-e was also involved in creating and distributing
%Locky \cite{cimpanu2021}. 
According to French court documents, Locky's developer is the same
individual who owned BTC-e, a fraudulent exchange~\cite{cimpanu2021}. Hence, the actor was able
to set up a new address for each payment without raising compliance alarms.
Locky is an early, less sophisticated example of a RaaS operation which would serve as an example to many cybercriminal actors to follow.

%\todo{What is its involvement with BTC-e telling us?
%Perhaps remove, or elaborate}.

\subsection{Ransomware Payment Characteristics}

\begin{figure}
	\includegraphics[width=\columnwidth]{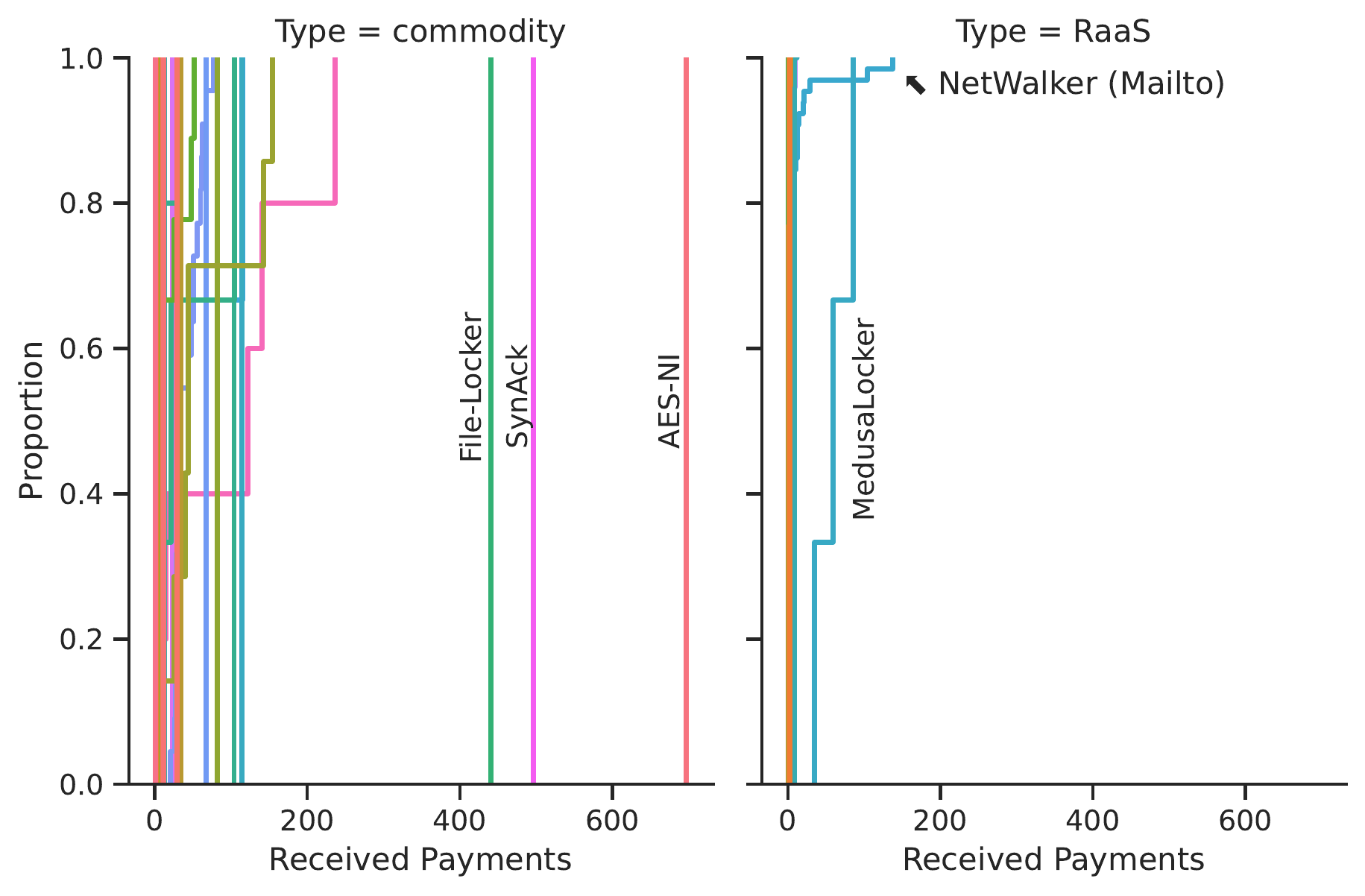}
	\caption{ECDF of payments per address for commodity ransomware and RaaS
actors.}
	\label{fig:ecdf-transactions}
	\end{figure}

RaaS actors are not only more effective in terms of profits, but also in handling payments. They typically have higher revenue per address,
while also generating unique addresses for victims. In
Figure~\ref{fig:ecdf-transactions} we show the cumulative distribution of received
payments between commodity and RaaS actors. Commodity
ransomware actors typically use single wallet addresses to receive
hundreds of ransom payments. The highest amount of payments to a single address
is 697 to AES-NI, followed by 496 to SynAck and 441 to File-Locker. While these
are outliers, Figure \ref{fig:ecdf-transactions} shows that using a single
address to receive upwards of 100 payments is not unusual.

In contrast, RaaS actors almost exclusively use a new wallet address to receive
each payment, as observed in Figure~\ref{fig:ecdf-transactions} (right). 
An outlier is an address associated with NetWalker
which has received 138 payments. This address is likely an intermediate payment
address, combining payments from many victims, discovered during McAfee Labs'
investigation into NetWalker~\cite{mcafee}. 

\begin{figure}
	\includegraphics[width=\columnwidth]{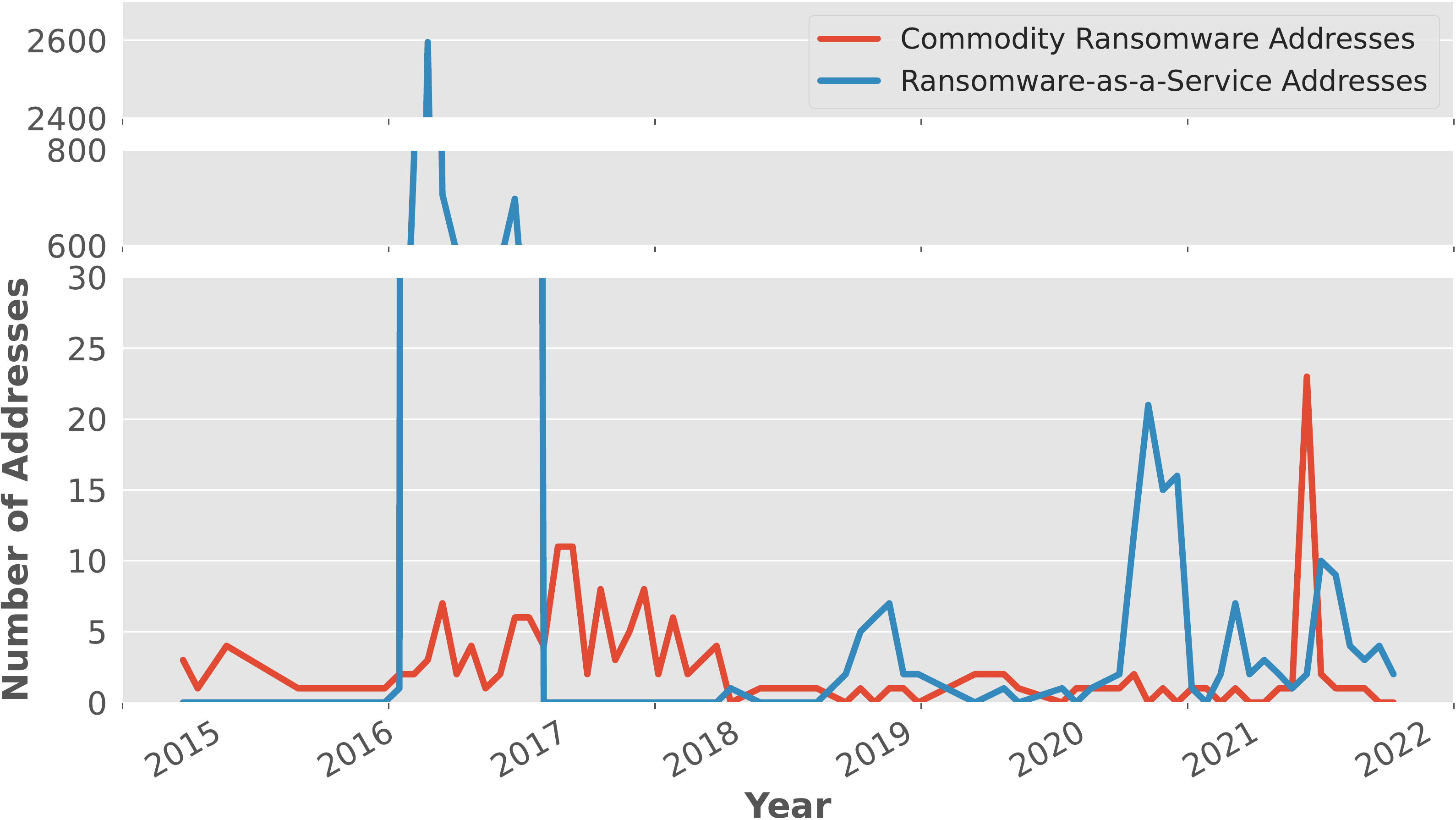}
	\centering
	\caption{Number of unique payment addresses for commodity ransomware and RaaS.}
	\label{fig:addressesperyear}
	\end{figure}

The distribution of unique addresses per commodity ransomware and RaaS 
actor is presented in Figure~\ref{fig:addressesperyear}. In stark contrast to
the revenue from ransom activities, presented in
Figure~\ref{fig:revenueperyear}, the number of addresses used in recent years
are low, on the order of tens per month.
We suspect that RaaS actors prefer
to create new addresses for each new ransom payment in order to ensure their
pseudo-anonymity, and thus make legal investigations and takedowns more difficult. 

%\todo{@George/Kris: discuss to remove Figure 6 as it is a bit redundant with the ECDF?}

%{fig:addressesperyear}

%families
%belonging to either the RaaS category or commodity group. The year is based on
%the timestamp of the first received/incoming (=victim) transaction.

%\begin{figure}
%	\includegraphics[width=3.5in]{figures/transactions_per_year.pdf}
%	\caption{Incoming Payments for Commodity and RaaS.}
%	\label{fig:transactionsperyear}
%	\end{figure}

%Figure \ref{fig:transactionsperyear} shows incoming transactions for RaaS and commodity families per year.

%Pointers from George:
%How many transactions are in BTC or in other cryptocurrencies. - begin of chapter
%Total number of instances for RaaS and RW
%Total number of transactions for RaaS and RW
%Total amount of BTC for RaaS and RW

%\begin{figure}
%	\includegraphics[width=3.5in]{figures/ecdf_number_transactions.pdf}
%	\caption{ECDFs of number of incoming transactions for commodity and RaaS
%addresses.}
%	\label{fig:ecdftransactions}
%	\end{figure}

%Figure \ref{fig:ecdf-transactions} shows the distribution of the number of
%incoming payments per address of both RaaS and commodity addresses. 
%Figure \ref{fig:addressesperyear} shows the unique addresses per commodity ransomware
%or RaaS group.

%Figure \ref{fig:ecdfreceived} shows received Bitcoin distribution of both RaaS and Commodity addresses. The Bitcoin blockchain accounts transactions using satoshis, which are 100th millions of one Bitcoin, so this is converted /1e8.

Moreover, our analysis shows that RaaS groups apply better operational security
practices when using native Bitcoin functionality for wallets (payment
addresses). Bitcoin uses Bitcoin Script to handle transactions between
addresses. The script type used defines the wallet type.
Pay-to-Public-Key-Hash (P2PKH) addresses have the prefix \textit{1}. This is
Bitcoin's legacy address format and the most common address format in our
dataset with 7,339 addresses. 46 addresses in our dataset are Pay-to-Script-Hash
(P2SH) formatted, recognized by the prefix \textit{3}. To spend received
payments in Bitcoin, the recipient must specify a redeem script matching the
hash. The script can contain functionalities to increase security, such as
time-locks or requiring co-signatures. We only observe this for select actors in
our dataset: Qlocker, Netwalker, REvil, Ryuk and Phobos. This could mean that
these groups have a specific interest in operational security, as transactions
usually aren't supported by exchange platforms. Another address format is
Pay-to-Witness-Public-Key-Hash (P2WPKH), or Segregated Witness (SegWit)
protocols, with prefix \textit{bc1q}. 72 addresses in our dataset, belonging to
Conti, Netwalker, SunCrypt, DarkSide and HelloKitty. These are RaaS actors and could imply that they deliberately use SegWit for additional
security instead of a traditional address format.

%SamSam indictment: https://www.justice.gov/opa/pr/two-iranian-men-indicted-deploying-ransomware-extort-hospitals-municipalities-and-public
%NetWalker indictment: https://www.justice.gov/opa/pr/department-justice-launches-global-action-against-netwalker-ransomware
%Seized REvil coins: %https://www.bleepingcomputer.com/news/security/fbi-seized-23m-from-affiliate-of-revil-gandcrab-ransomware-gangs/
%DarkSide coins seized: https://www.justice.gov/opa/pr/department-justice-seizes-23-million-cryptocurrency-paid-ransomware-extortionists-darkside

\section{Money Laundering Analysis}\label{sec:laundering} %\todo{@George: confirm flow; 1.75 pages}

In the previous section, we investigated ransom payments by victims to ransomware
actors. In the next section, we investigate 13,097 laundering transactions in
our dataset (see Table~\ref{table:datasetoverview}) to shed
light on how these actors liquidate their illicit earnings.

\subsection{Laundering Strategies}

\begin{figure}[t]
	\includegraphics[width=\columnwidth]{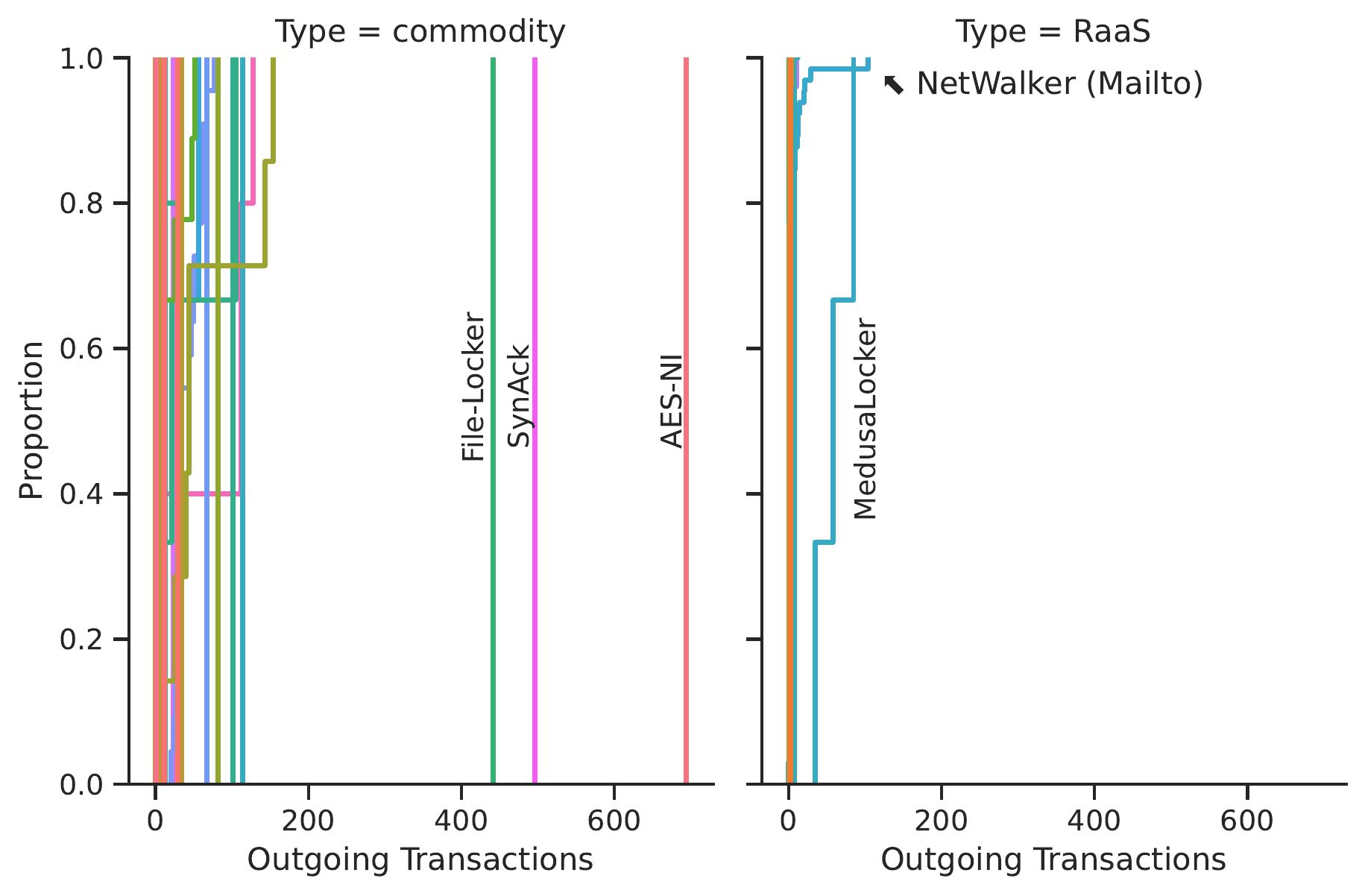}
	\caption{Transfer transactions per Address for commodity and RaaS actors.}
	\label{fig:spentoutputs}
\end{figure}

To avoid exposing their identity, ransomware actors will usually launder their
revenue. After routing funds through one or more services to obfuscate the money
trail, it is cashed out as legal tender or monetized through the purchase of
voucher codes or physical goods.
In Figure \ref{fig:spentoutputs} we show the number of transfer transactions per address. 
%This is opposite to Figure \ref{fig:ecdf-transactions} with incoming payments, discussed earlier. Both data points are stored on the Bitcoin blockchain. 
The number of transfer (outgoing) transactions provides insights on how actors
prefer to initialize their laundering. In short, we see that RaaS actors mostly
prefer to empty the deposit address in one transaction, whereas commodity actors
prefer multiple smaller transactions, up to hundreds, in some cases more.
Hence commodity ransomware actors are less sophisticated. For example, three
commodity ransomware actors with the most payments per address
(File-Locker, SynAck, AES-NI) also have the most outgoing transactions. While
the motivation for this behavior remains unclear, given that law
enforcement scrutiny was relatively low, it is likely that the commodity actors
took advantage of the ability to cash out more frequently with little risk. This
is further supported by their choice of laundering entities.

\begin{figure}[t]
	\includegraphics[width=\columnwidth]{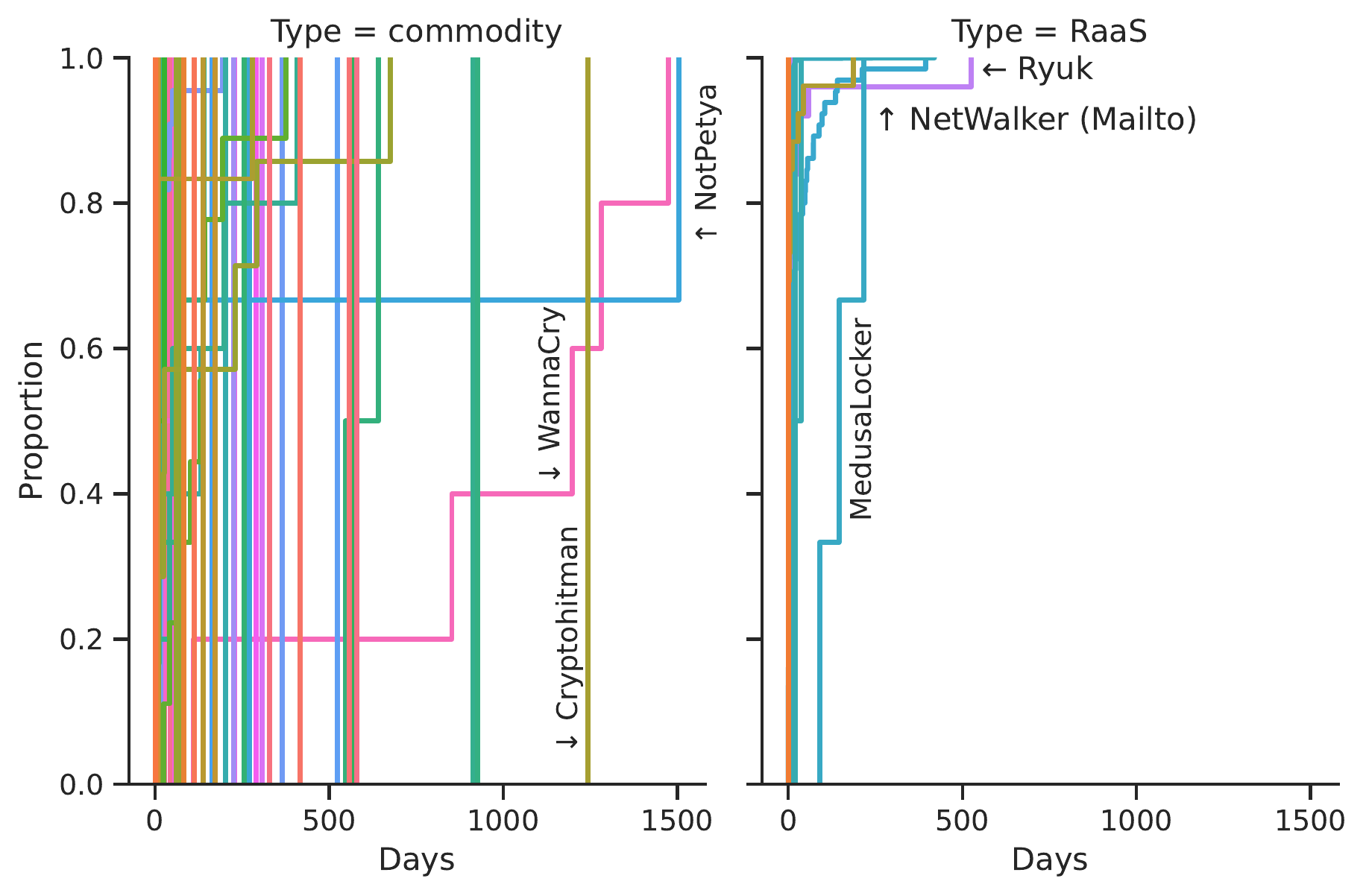}
	\caption{Collect-to-Laundry time for commodity ransomware and RaaS actors.} % to keep received payments on deposit address.}
	\label{fig:timepreference}
\end{figure}

\begin{table*}[!bpt]
	\centering

	\caption{Laundering Entities Overview}
	\scriptsize
	%\scalebox{0.55}{
	\begin{tabular*}{\linewidth}{lll}
		\hline
		\textbf{Entity} 				& \textbf{Description} 										& \textbf{Evidence}		\\ \hline
		ATM / Payment Provider			& Payment gateways for physical/online merchants or ATMs, usually used to launder small amounts     		& \cite{justice2021}	\\ \hline
		Dark Market / Illegal Services	& Illegal services available on Tor or other Internet services, used to buy illegal server hosting etc.		& \cite{europol2021}    \\ \hline
		Fraudulent Exchange				& 
		Exchange platforms officially sanctioned by the US Office of Foreign Assets Control (OFAC)          & \cite{cimpanu2021} 	\\ \hline           
		Gambling						& Online casinos and gambling platforms, used to launder small amounts anonymously            			& \cite{ft2021}			\\ \hline
		Low/Moderate ML-Risk Exchange	& Exchanges with strict AML/KYC policies might still be used for laundering criminal funds			& \cite{binance2022}	\\ \hline
		Mixers							& These services take and `mix' Bitcoin from various parties to obfuscate ownership     	& \cite{justice2021a}	\\ \hline
		(Very) High ML-Risk Exchange	& Exchanges with lax or no AML/KYC implementations are popular for money laundering			& \cite{wsj2021}        \\ \hline
		Wallet Service					& Custodial/online wallets, some might have also have privacy features such as mixers.	& \cite{ft2021}         				\\ \hline 
	\end{tabular*}
	%} %end scalebox
	\label{table:entities}
\end{table*}

Almost all ransomware actors in our dataset launder their proceedings entirely.
The speed with which this happens can be inferred from the time between the
first incoming payment to and the last outgoing transaction from the deposit
address. We define this time duration in which ransomware actors start laundering
after having received the payment as \textit{collect-to-laundry time}. Note that
this is not the total duration for caching the ransom, but rather the
time spent between start receiving the ransom payment and transferring the payment received. 
Figure~\ref{fig:timepreference} shows the ECDF of the collect-to-laundry time
(in days) for the commodity ransomware and the RaaS actors in our
dataset. 
%The time span in days is between receiving the first payment and the last
%recorded outgoing transaction shows how quick actors are to launder ransom
%payments. It can be observed that 
RaaS actors have a significantly lower collect-to-laundry time
compared to commodity actors. Typically, payments to RaaS actors are transferred away from the
deposit address in the first minutes to hours after payment. The few outliers in
RaaS are caused by NetWalker and individual addresses associated with actors for
which we have multiple addresses in our dataset (Ryuk, Conti). As the illicit
funds received by RaaS are washed out quickly and, typically, in full, this suggests that it is more difficult to
track payments to RaaS and lowering the odds of recovery.

Only a small set of families still have significant portions of their
proceedings on the original address. This is the case for
NetWalker, which has 20.36\% still on an address, MedusaLocker (7.98\%) and
WannaCry (7.92\%). In this case, it is likely that the actor has lost the
private key or is incapable to safely launder the ransom, for example due to law
enforcement scrutiny. It is known that NetWalker's proceedings have been seized by
law enforcement \cite{Netwalker}, with WannaCry under heavy monitoring and
most of the laundering failed \cite{Wannacry}.

\subsection{Challenges in Fighting Laundering}

\begin{figure}[t]
	\includegraphics[width=.9\columnwidth]{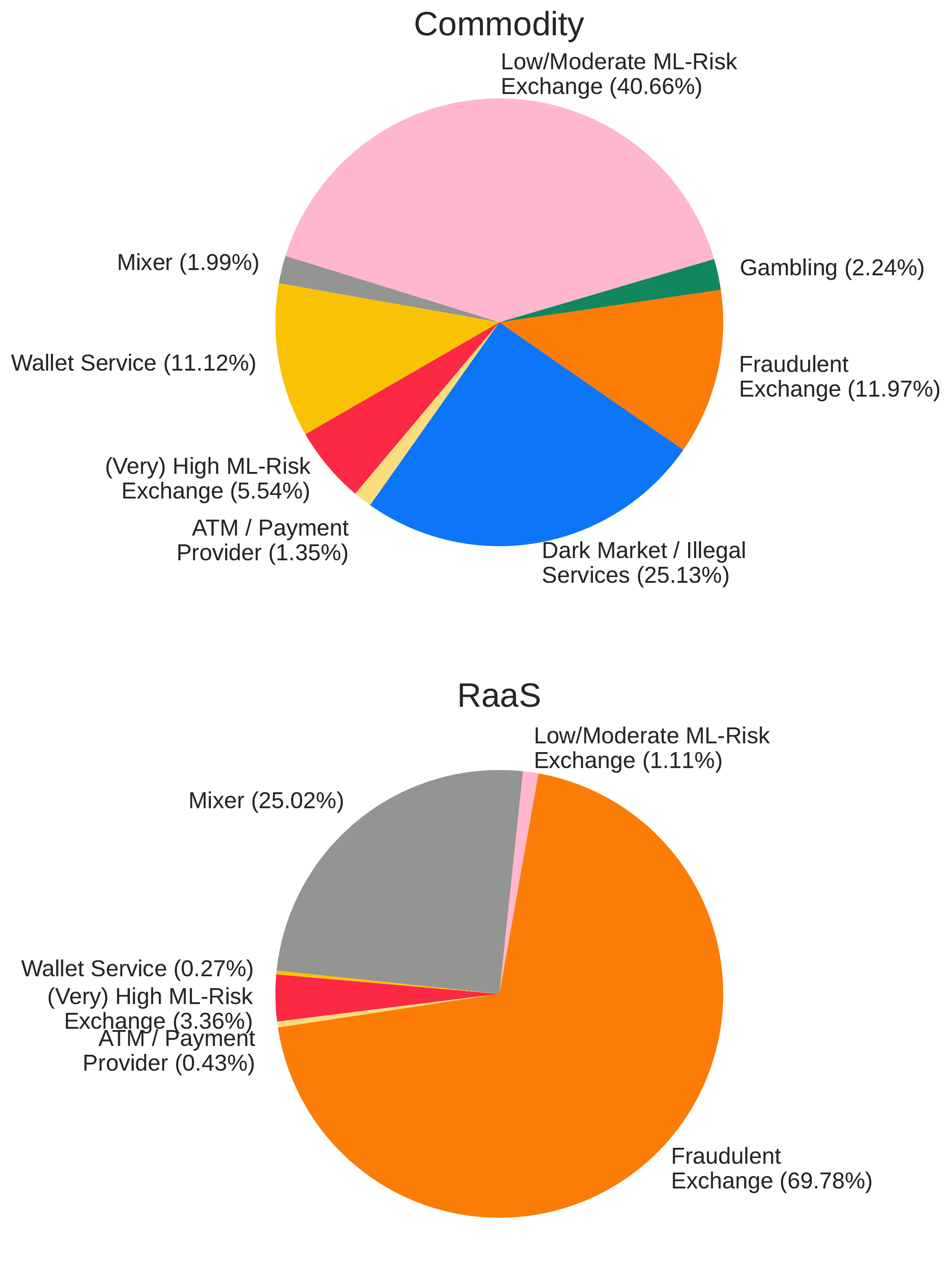}
	\caption{Pie chart of one-hop laundering entities.}
	\label{fig:laundering}
\end{figure}

Contrary to popular belief, Bitcoin isn't anonymous but pseudo-anonymous. Forensic
analysis might link a Bitcoin address to a real-world identity, especially when
an exchange platform is used to convert between fiat currency and Bitcoin. In
most jurisdictions, legal entities behind such platforms are held to Know Your
Customer (KYC) legislation, which requires them to verify the identity of every
user signing up on their service. During an investigation, when known illicit
Bitcoin is routed through an exchange that requires KYC, authorities have a
chance to identify the culprit. Several industry players support law enforcement
in such AML investigations, with technology based on clustering algorithms which
can link addresses to a service such as an exchange platform. As seen in Figure~\ref{fig:laundering}, we have grouped the data we obtained through Crystal
Blockchain in a select set of entities, which are explained in Table~\ref{table:entities}.

Laundering can involve routing illicit funds through several hops before cashing
out. As it is difficult to know where actual ownership has terminated after
several hops, in this analysis we only regard the first hop, i.e., the first
transfer transaction. This is the service
to which actors transfer funds directly after having obtained them at the deposit address
shared with the victim. As this has the closest link to the payment address,
this is the first point of investigation for law enforcement. An actor choosing to use a service implies that they trust the service, at least
enough not to disclose their identity. 

Figure~\ref{fig:laundering} shows the proportion of estimated USD value of Bitcoin directly transferred (first hop) to the entities explained in Table~\ref{table:entities} for commodity and RaaS actors. Due to limitations in reliably establishing (legal) entities behind an address, the direct transactions in our dataset account for a subset of the total revenue generated by the actors in our dataset. Hence we report using percentages, a best practice used with comparable datasets \cite{wang2021large}.

Our core observation is that commodity actors don't exhibit a specific laundering strategy, while
RaaS actors primarily use fraudulent exchanges and mixers. Mixers are services which take in Bitcoin from cybercriminals or simply privacy-aware users and combine these in many transactions. Through this the
accurate tracking of Bitcoin is hindered, as every client gets their initial
deposit (minus service fee) back as a mix from other users' Bitcoin.
Thus, it is more difficult to trace the laundering activity of RaaS criminal
actors.

When considering fraudulent exchanges together with low- and high-risk
exchanges, commodity authors seem to prefer exchanges, and thus perhaps cash-out
to fiat currency or other cryptocurrencies. It is however also known that
cybercriminals have wound down the use of fraudulent exchanges~\cite{oosthoek2020cyber}. In a sense, commodity actors
do not partake in any systematic laundering at all, whereas RaaS actors use fraudulent (non-KYC) exchanges and mixers,
a clear laundering strategy. Based on this, we hypothesize that the chances of recovering payments through law enforcement intervention are
higher with commodity ransomware than with RaaS. The services of their choice logically leave more user traces (IP address, login session) than mixer services and fraudulent exchanges designed to obfuscate ownership.

When an actor's collect-to-laundry time is high, a law enforcement
investigation may be able to successfully recover the funds.  However, in many such cases there is
less incentive to intercept transactions due to the comparatively low ransom
amounts. The speed with which RaaS groups transfer funds out suggests criminal sophistication, which is also
reflected in their preferred means of laundering. Given this, it is difficult to intercept funds unless law
enforcement is already involved at the very moment the payment is
made~\cite{darkside2}.

\section{Conclusion}\label{sec:conclusion} %\todo{@George: do a pass}

Ransomware is a severe, growing threat plaguing our world. In this paper, we
take a data-driven and ``follow the money'' approach to characterize the
structure and evolution of the ransomware ecosystem. To this end, we report on
our experience in operating {\tt Ransomwhere}, our open crowdsourced ransomware payment tracker to collect information from victims of ransomware attacks.
 By analyzing a corpus of
more than 13.5k ransom payments with a total revenue of more
than \$101 million, we shed light on the practices of these criminal actors over
the last years. Our analysis unveils that there are two symbiotic, parallel markets. Commodity ransomware actors, and 
(dominant since 2019) Ransomware as a Service (RaaS) actors. The
first is operated by individuals or a small group of programmers, the second by
professional criminals who offer it on an affiliate basis to typically less technical criminal actors.
Due to differences in victimization, the first has low ransom amounts, the
latter higher ransom amounts depending on the victim profile.

Our analysis of ransom payments (all in Bitcoin) shows that RaaS actors have
adopted more sophisticated cryptographic techniques, compared to commodity
actors, in their operation and typically generate one address per victim to
hide their identity. This allows RaaS to generate more revenue and with
higher level of protection, attracting more criminal groups
to use RaaS to perform high profile attacks in recent years.
RaaS actors are also more efficient to launder
ransom payments, as they move to launder funds within hours
or days. RaaS actors also transfer revenue from ransom payments to mixers and other
sophisticated laundry entities that make difficult for law enforcement agencies
to recover ransom payments.

%Ransomware has enabled a lucrative cyber-criminal business model, monetizing the
%poor state of IT security in many organizations. We have analyzed more than 12k
%ransom payments accounting for 101 million USD in total. Attributing this as
%economic damage due to technical debt caused by old legacy vulnerable
%Internet-connected device in the victim organizations, is taking a one-sided
%view on a complex problem area. Bitcoin might have made cyber-criminal
%undertakings easier, but malware with credit card payment functionality existed
%already before Bitcoin.
%
%Exchanges and other services with weak or no checks and balances against
%money-laundering are the true enabler. In our analysis we have found that
%ransomware actors almost exclusively use such services as it obfuscates
%identification and leads forensic analysis of money flow astray. The rising
%proceedings from these attacks provide ransomware actors with more budget than
%the average organization's IT security budget.
%
%In this paper, we have followed the money. In order to defend against ransomware
%for the years to come, \textit{follow the money} should become a central axiom
%to ransomware defense: to understand the actors, their incentives and to
%potentially recover ransom payments.

%\subsection*{Acknowledgements}

\balance
\bibliographystyle{ACM-Reference-Format}
\bibliography{paper}

%\appendix
%\input{sections/traffic_classes}

\end{document}